# The infrared imaging spectrograph (IRIS) for TMT: spectrograph design

Anna M. Moore*[a], Brian J. Bauman[b], Elizabeth J. Barton[c], David Crampton[d,e], Alex Delacroix[a], James E. Larkin[f], Luc Simard[d,e], Ryuji Suzuki[d,g] and, Shelley A. Wright[h]

[a]Caltech Optical Observatories, 1200 E California Blvd, Pasadena, CA, USA 91125;
[b]Lawrence Livermore National Lab., 7000 East Avenue, Livermore, CA USA 94550;
[c]Univ. of California, Irvine, CA, USA, 92697;
[d]Thirty-Meter Telescope Project, 2632 E. Washington Blvd., Pasadena, CA, USA 91107;
[e]National Research Council of Canada, Herzberg Institute of Astrophysics, 5071 West Saanich Rd., Victoria, Canada, V9E 2E7;
[f]Univ. of California, Los Angeles, 405 Hilgard Avenue, Los Angeles, CA USA, 90095;
[g]Subaru Telescope, 650 North A'ohoku Place, Hilo, HI USA, 96720;
[h]Univ. of California, Astronomy Department, 601 Campbell Hall, Berkeley, CA, USA, 94720.

## ABSTRACT

The Infra-Red Imaging Spectrograph (IRIS) is one of the three first light instruments for the Thirty Meter Telescope (TMT) and is the only one to directly sample the diffraction limit. The instrument consists of a parallel imager and off-axis Integral Field Spectrograph (IFS) for optimum use of the near infrared (0.84um-2.4um) Adaptive Optics corrected focal surface. We present an overview of the IRIS spectrograph that is designed to probe a range of scientific targets from the dynamics and morphology of high-z galaxies to studying the atmospheres and surfaces of solar system objects, the latter requiring a narrow field and high Strehl performance. The IRIS spectrograph is a hybrid system consisting of two state of the art IFS technologies providing four plate scales (4mas, 9mas, 25mas, 50mas spaxel sizes). We present the design of the unique hybrid system that combines the power of a lenslet spectrograph and image slicer spectrograph in a configuration where major hardware is shared. The result is a powerful yet economical solution to what would otherwise require two separate 30m-class instruments.

## 1. INTRODUCTION

### 1.1 Science case

The IRIS IFS science case has been developed over several years and is both varied and impressive. A 30m-class telescope operating close to the diffraction limit is an extremely powerful device capable of achieving a revolutionary advancement over current observations. Several science cases benefit from the diffraction limited imager in addition to the IFS. A comprehensive discussion of the IRIS science case can be found elsewhere ([1], [2], [3]) and we present an appropriately condensed version here. The modeled performance of the IRIS imager and IFS is also presented at this meeting ([4]).

The IRIS IFS will:

- Study solar system atmospheres and surfaces, like Titan, Io and Kuiper Belt objects;
- Characterize exoplanets, initially detected by other instruments, that currently do not have the collecting aperture to acquire spectra of these tiny, faint objects;
- Categorize self-luminous exoplanets at large separations;
- Identify individual stars in the central crowded regions of M31, M32 and M33 and categorize their stellar content.

*amoore@astro.caltech.edu; phone 1 626 378-4291; fax 1 626 568-1517.

- Calculate the masses of black holes in distant galaxies far outside the realm of what is currently achieved;
- Perform a spatial dissection of galaxies during the peak epoch of galaxy formation in the range z~1-4, the most active known period of star formation and AGN accretion in the history of the Universe;

**1.2 Specifications**

The top-level specifications of the spectrograph are dictated by the science requirements and influenced by technical considerations and are summarized in Table 1. The resulting observing modes for the IRIS IFS are presented in Section 8.

Table 1: Top level specifications for the IRIS IFS

|  | FOV (arcsec) | Format on sky (pixels/spaxels) | Bandpass | Resolution | Sky and Detector sampling |
|---|---|---|---|---|---|
| Spectrograph Channel 1 (4mas) | 0.45 x 0.64 | 112 x 128 (14,336) | 5% max | 4,000 8,000 10,000 | 4 mas/ lenslet 2 pixels/ lenslet pupil image |
| Spectrograph Channel 2 (9mas) | 1.01 x 1.15 | 112 x 128 (14,336) | 5% max | 4,000 8,000 10,000 | 9 mas/ lenslet 2 pixels/ lenslet pupil image |
| Spectrograph Channel 3 (25mas) | 1.125 x 2.20 | 45 x 88 (3960) | 20% max | 4000 8000 | 25 mas/ slicer 12.5mas per pixel No anamorphic mag |
| Spectrograph Channel 4 (50mas) | 2.25 x 4.40 | 45 x 88 (3960) | 20% max | 4000 8000 | 50 mas/ slicer 25mas per pixel No anamorphic mag |

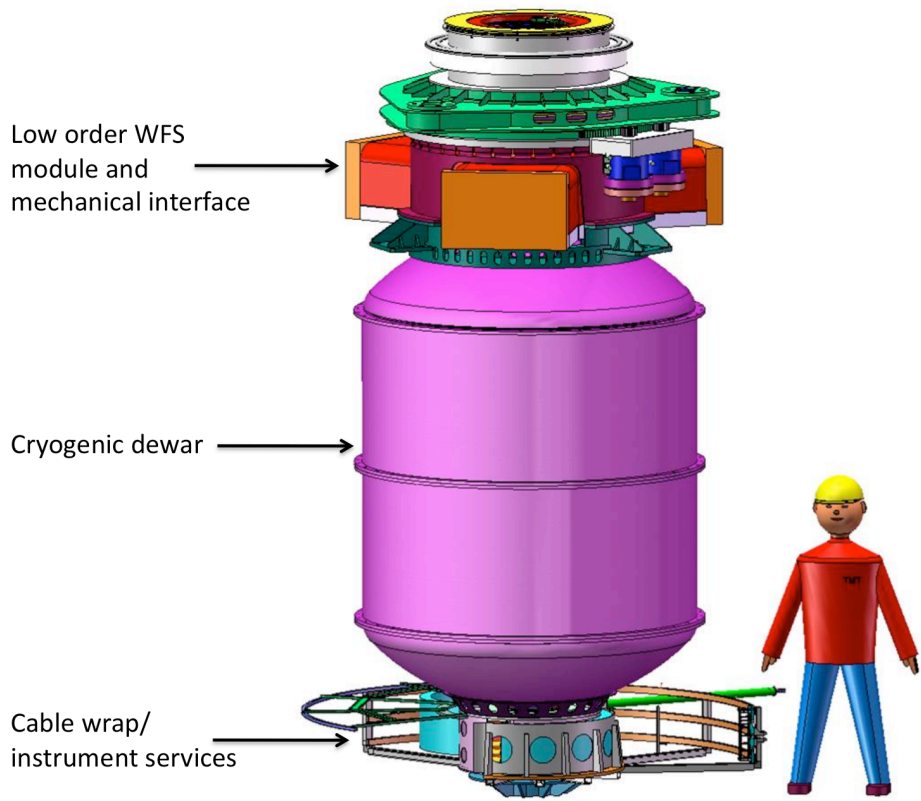

Figure 1: The IRIS instrument consists of a large cryogenic dewar, a cooled low order WFS module (OIWFS) and a cable wrap providing essential services to the instrument. A 1.7m Solidworks figure is provided for size reference.

## 2. IRIS LAYOUT

### 2.1 IRIS Layout

A rendering of the IRIS instrument is shown in solid view in Figure 1 and sectioned in Figure 2 [*left*]. The instrument consists of a large cryogenic dewar 2m diameter by 3.25m in length, mounted to a cooled chamber containing three low order wavefront sensors, the On-Instrument Wavefront Sensor module [8]. IRIS is mounted and rotated in a vertical orientation therefore is gravity invariant as shown in Figure 2 [*right*]. The cryogenic dewar houses the IRIS imager, Suzuki et al [5] and spectrograph. The imager and spectrograph are operated in parallel so utilizing as much of the precious 30m telescope focal plane as possible. Further details of the instrument mechanical layout and interface to the NFIRAOS Adaptive Optics system [6] for TMT can be found in Larkin et al [7].

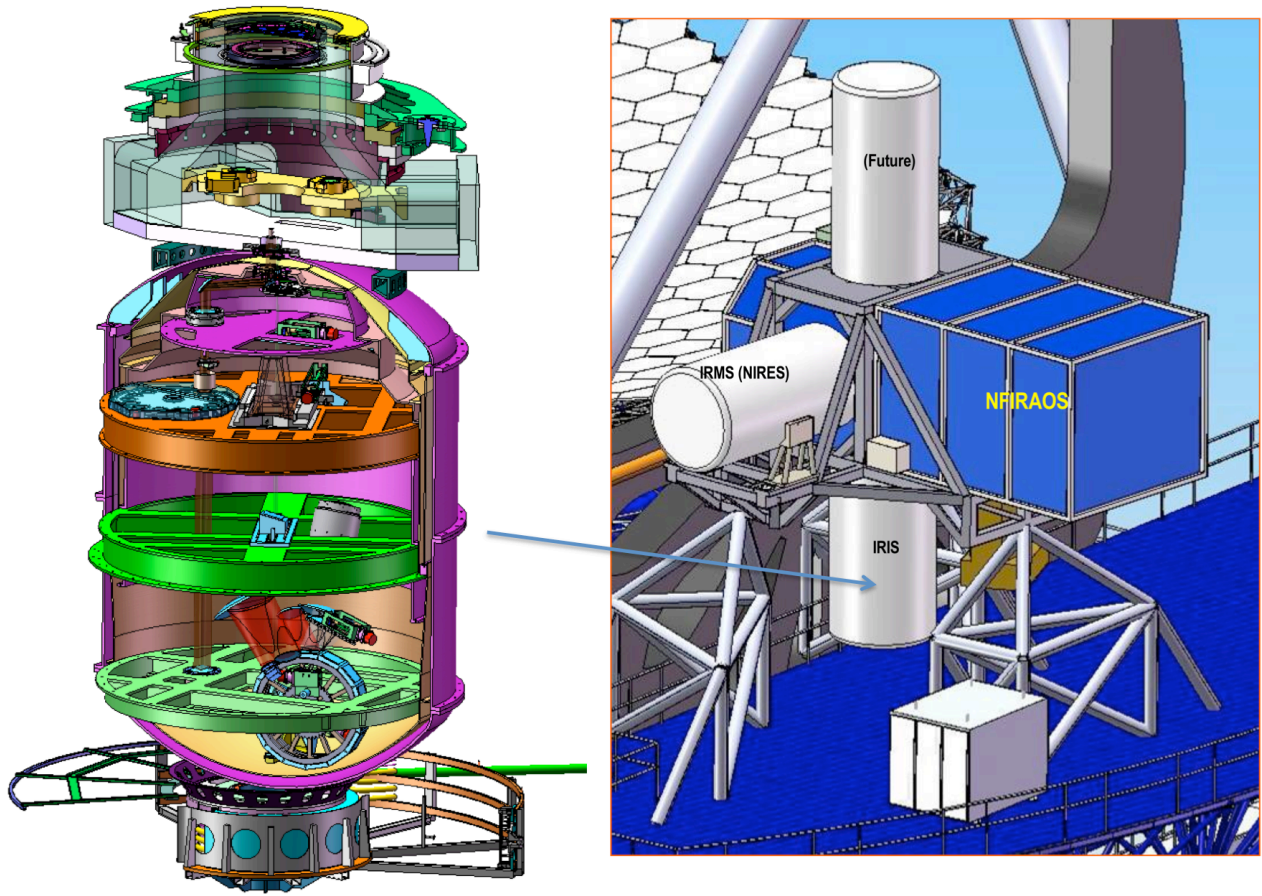

Figure 2: [*left*] A section view of IRIS showing the optical channels including a 17" diffraction limited imager and 4-channel IFS spectrograph; [*right*] IRIS mated to the TMT Adaptive Optics system NFIRAOS.

# 3. THE HYBRID SPECTROGRAPH

## 3.1 Results from IRIS Feasibility study

During the IRIS feasibility phase (2004-2005) a trade study was performed between a lenslet-only and image slicer-only integral field spectrograph for IRIS. The conclusion reached was that each technology had its advantages and disadvantages. Table 2 summarizes these conclusions. The results for each of the four plate scales dependent upon the technology used is shown along with a brief description of the pros and cons of each method. Both cases use the same number of spaxels as in the present design (112x128 lenslets and 88 slicers with 45 spaxals each).

The finest spaxel scale, 4mas, samples the diffraction limit of the telescope. The IFS must have an extremely low residual wavefront error (WFE<30nm) for this to be achieved in practice. We believe the lenslet array is superior at the finest scale since it samples directly the PSF and has inherently excellent image quality. In addition, the field of view is sufficient to image a large fraction of a PSF halo. At the coarsest scale, the slicer is superior since the lenslet would be wasting most of its spaxels on sky and can't be reformatted to satisfy the larger required bandpass. The two intermediate scales are more ambiguous but the FOV benefit of the lenslet aids with the 9 mas scale while the slicer makes better use of its pixels in the 25 mas scale. What the tables do not show is the difficulty that both spectrograph styles have to handle the full range of platescale (factor of 12.5) without a complete change of the lenslet or slicer unit. Each can, however, accomplish 2 neighboring platescales (roughly factor of 2) with only minor modifications.

Table 2: Results of image slicer-only versus lenslet-only technology for IRIS that led to the choice of the hybrid design.

**Hypothetical Parameters for Slicer-only Spectrograph (shaded=poor performance, white=good performance)**

| Scale (mas/spaxel) | FOV (") | Bandwidth @ R=4000 | Comment |
|---|---|---|---|
| 4 | 0.18x0.35 | 20% | Poor wavefront quality, very small field |
| 9 | 0.40x0.79 | 20% | Image acceptable, small field |
| 25 | 1.13x2.20 | 20% | Field acceptable, good use of pixels for bandwidth |
| 50 | 2.25x4.40 | 20% | Good FOV, good bandwidth |

**Hypothetical Parameters for Lenslet-only Spectrograph**

| Scale (mas/spaxal) | FOV (") | Bandwidth @ R=4000 | Comment |
|---|---|---|---|
| 4 | 0.45x0.51 | 5% | Reasonable field for small scale, excellent image quality |
| 9 | 1.01x1.15 | 5% | Good FOV, good images |
| 25 | 2.80x3.20 | 5% | FOV more than sometimes needed, bandwidth becoming drawback |
| 50 | 5.60x6.40 | 5% | FOV excessive, strongly prefer bandwidth |

## 3.2 The Hybrid design

We will show that a fine-scale lenslet-based spectrograph and a coarser-scale slicer-based spectrograph can share gratings, camera optics and IR detector. This means that although IRIS has 2 slicing methods and four plate scales there can be a direct sharing of the most expensive hardware, namely the cryogenic grating turret, the 4k x 4k detector and associated hardware and the spectrograph camera that is a large three-mirror anastigmat (TMA).

**Matching camera F ratio**

The predominant feature that permits the hybrid design is the similarity of the required camera focal ratio of the largest spaxel scales of the lenslet and slicer IFS, ~F/4, assuming a Nyquist sampled image is required to be sampled by 15μm pixels at the shared 4k x 4k detector. The respective collimators, that must be different, produce a 100mm diameter pupil for the largest plate scales at the grating location. This permits a sharing of the grating turret and many of the individual gratings housed on the turret. The remaining task is to produce a feasible layout. To achieve this we position the lenslet IFS to be in straight-through mode, as shown in Figure 5 [*right*] while the slicer is selected by moving a fold mirror into the beam. It was found that the image slicer IFS, with a simple refractive collimator, was more flexible to geometry changes.

The remaining sections are ordered as follows. The method of selection between the 4 available plate scales is summarized in Section 4. Each of the four channels requires a set of pre-optics to demagnify the native F/15 feed from TMT to the lenslet array or image slicer stack. The four sets of pre-optics are discussed in Section 5. The lenslet IFS channel is described in Section 6 and corresponding image slicer IFS in Section 7. The filter and gratings set for the spectrograph system are summarized in Section 8. The IRIS spectrograph mechanical design is addressed briefly in Section 9.

## 4. IFS CHANNELS

IRIS offers four spaxel scales (4mas, 9mas, 25mas and 50mas) with only one of the four selected at any one time. IRIS offers parallel imaging and spectroscopy – the 17"x17" imager field is offset from the smaller IFS field.

The IFS channel of choice is selected by activating either 1 or 2 cryogenic stage mechanisms inside the IRIS dewar. The selection method is outlined in Figure 4 (slicer, 25/50mas, [*left*]) and Figure 5 (lenslet, 4/9mas [*left*]). The mechanical layout of the IRIS IFS is shown in both figures with the respective optical beams shown for the largest spaxel scale.

## 5. PRE-OPTICS

### 5.1 Common collimator

Light from NFIRAOS first encounters a refractive collimator once through the IRIS dewar entrance window. The collimator, shown in Figure 3 [*left*], has a 100mm focal length and produces a 6.67mm pupil at which the cold stop is placed. This collimator is common to all the spectrograph and imager optics, so was designed as a stand-alone component. A two-glass achromat suffices and spot diagrams for the design of choice are shown in Figure 3 [*right*].

The doublet is air-spaced to avoid any non-matching thermal coefficient of expansion issues. There could be a slight transmission improvement (perhaps 1%) throughout most of the science bands achieved by cementing elements and eliminating two anti-reflective coatings, but even if the glasses were chosen to minimize CTE effects, gains could well be offset by cement transmission losses that tend to occur >2.2μm. The next design phase should give at least some consideration to one-glass and three-glass designs to optimize overall cost/design performance.

A reflective design was not seriously considered because it produces the wrong packaging feature (it sends the beam back towards the input of the instrument, which is very nearby, and because of the relatively large field that the collimator would need to handle, compared to the focal length (a field of 10x10arcsecs is ±15.4 mm along the diagonal, or ±8.8° field of view for a 100 mm focal length—much too large).

Because of the relatively small thicknesses of the lenses required, the glass choices were somewhat broader for the collimator than they might be otherwise. A BaF2/S-TIH11 combination performs well. S-TIH11 would probably not be acceptable in thicknesses of 25 mm (88% throughput), but for 3 mm, the losses are acceptable. The doublet consists of a biconvex BaF2 element and a negative meniscus S-TIH11 element.

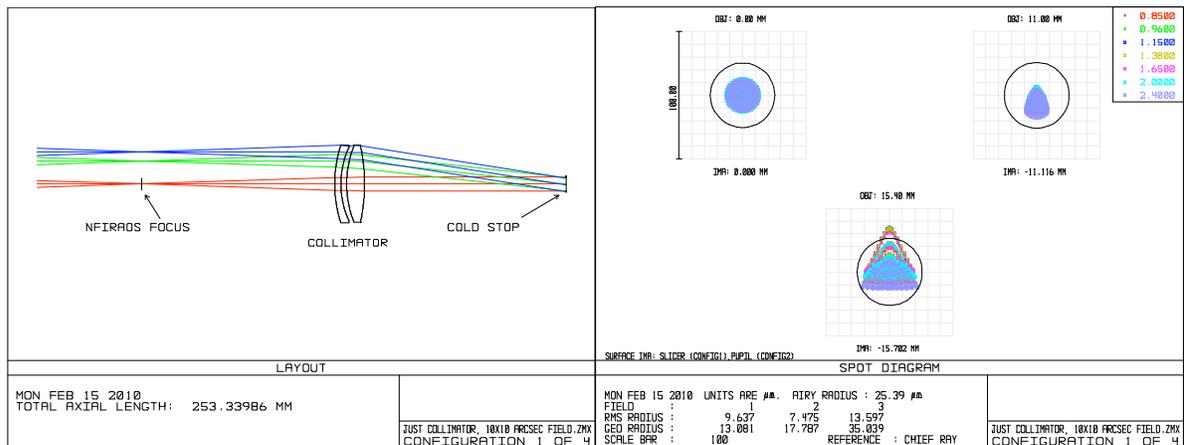

Figure 3: [*left*] Layout of pre-optics collimator at entrance to instrument, leading to the cold stop. The elements are Ohara S-TIH11 and BaF$_2$. [*right*] associated spot diagrams.

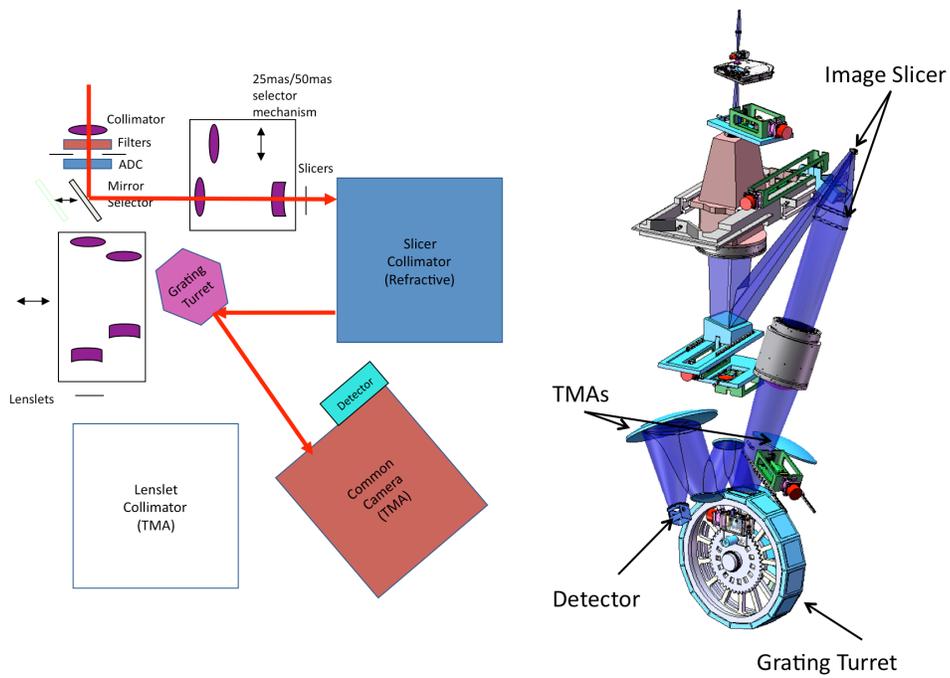

Figure 4: The 50mas and 25mas spaxel scales are the image slicer IFS channels. A schematic of the optical components utilized for these channels is shown [*left*]. A rendered drawing of the optical path [*right*] is shown for the 50mas channel with major hardware items identified.

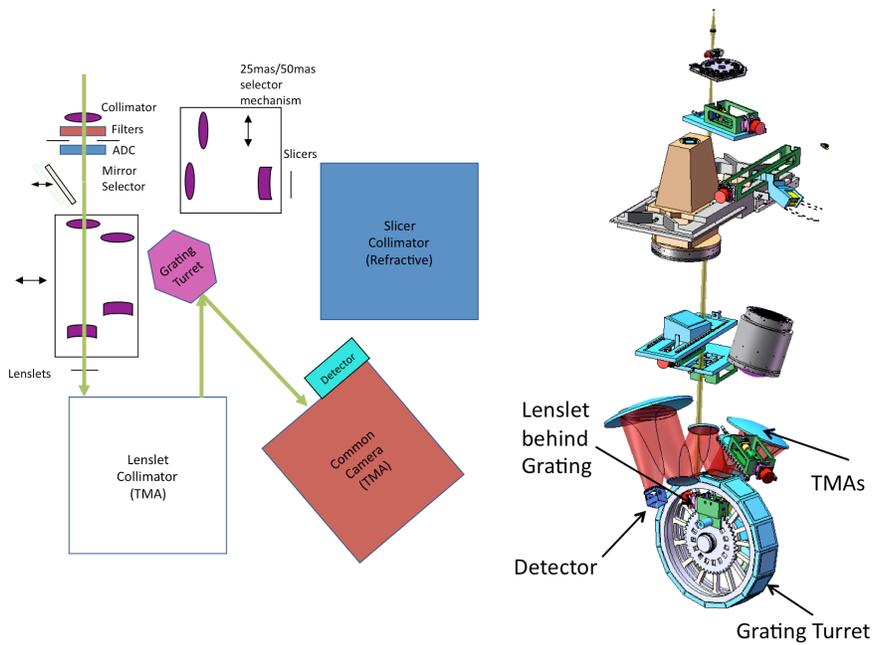

Figure 5: The 4mas and 9mas spaxel scales are the lenslet IFS channels. A schematic of the optical components utilized for these channels is shown [*left*]. A rendered drawing of the optical path [*right*] is shown for the 9mas channel with major hardware items identified.

## 5.2 Multiple camera optics

Following the refractive collimator and cold stop, there are two sets of pre-optics for the lenslet spectrograph, and two sets for the slicer spectrograph. The two different sets yield different plate scales but are otherwise interchangeable, with the same image plane and exit pupil position. First-order optics requires two separated optics in order to achieve the desired f#, image position, and pupil position. Some of the optics are singlets, and others are achromatized with a second glass. The last surface in each set is aspherized in order to control pupil wander—one surface suffices. Because the beams are quite slow at the last surface, the image quality is not significantly impacted by the asphere and the surface can be designed to steer the chief rays appropriately. Top-level specifications for the IFS pre-optics are shown in Table 3, optical layouts are presented with corresponding spot diagrams in Figure 6.

Table 3: Specifications for the IRIS IFS pre-optics

|  | Slicer spectrograph pre-optics | Lenslet spectrograph pre-optics |
| --- | --- | --- |
| Field sizes | 2.2 x 4.4 arcsec and 1.1x2.2 arcsec | 1.152 arcsec square and 0.512 arcsec square |
| F#'s | f/55 (50 mas/pixel) and f/110 (25 mas/pixels) | f/272 (9 mas/pixel) and f/603 (4 mas/pixel) |
| Overall length (NFIRAOS focus to slicer/lenslets) | 2543 mm | 3108 mm |
| Exit pupil position | 328 mm beyond slicer | 91.7 mm in front of lenslets |
| L1 features | BaF2/S-NPH2 doublet, 30 mm diameter | BaF2/S-NPH2 doublet; S-NPH2 singlet |
| L2 features | BaF2/S-TIH11 doublet, 120 mm diameter, aspheric last surface | BaF2/S-NPH2 doublet; BaF2 singlet, aspheric last surface |

# 6. LENSLET IFS

## 6.1 Lenslet array

The lenslet array has a square pattern and a pitch of 350 microns per lenslet. Each lenslet has a focal length of 1.3mm and thickness of 1.5mm. The lenses are plano-convex. The array is placed in the focal plane produced by the lenslet pre-optics and the pupil produced by each lenslet is typically 32 times smaller than its own diameter including the effects of pupil diffraction. This compression is sufficient to allow 16 spectra to interleave together in the height of an individual lenslet. A basic pattern of 16x16 lenslets is used as a building block for the full field. A 7x8 pattern of these basic blocks is then used to create the 112x128 spatial samples. The spectrum from an individual lenslet can extend for the length of a 16x16 block or approximately 512 pixels. In the dominant modes where the spectral resolution is 4000, this corresponds to a 5% backpass in each individual exposure. Details of the spectral layout at the detector can be found in Larkin et al [7].

## 6.2 Lenslet collimator and camera

The lenslet collimator and the camera TMA are the most difficult items of the IFS optical design, so those subsystems were allowed to drive other subsystems. The lenslet collimator is fast since the lenslet f# is f/2.64 along the diagonal. An additional difficulty in that the field is also quite large. Details of the lenslet double TMA are shown in Figure 7.

A brief summary of the double TMA system is as follows:

- Both the lenslet collimator and camera TMA's consist of three aspheric mirrors, of $8^{th}$ order polynomials, with largest element ~1m in diameter;
- Initial designs used a common optical axis for each TMA, but it was found that performance improved by moving the aspheres nearest the grating modestly off-axis. The remaining two aspheres in each TMA share a common axis;
- The order is optimal – there is no benefit in trying to reduce the order according to the vendor approached;
- Alignment tolerances are achievable;
- The required stability tolerances are more demanding – of order $1/100^{th}$ the alignment tolerances. Prototyping will almost certainly be required during the next phase to investigate this issue further.

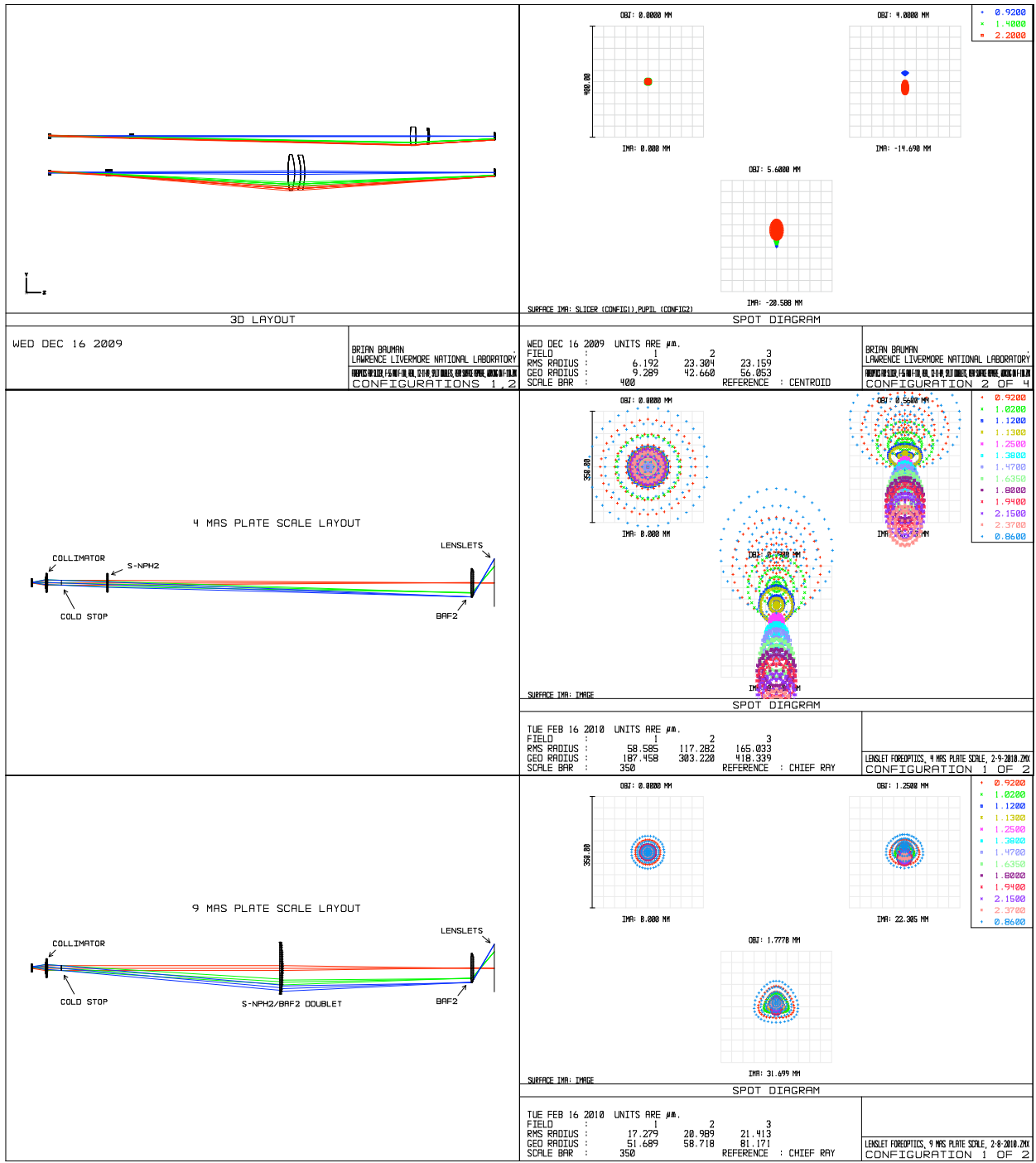

Figure 6: [*upper left*] layout of 25 mas and 50 mas plate scale slicer pre-optics. [*upper right*] spot diagrams for various points in the field for the 50mas spaxel scale. The grid is the width of a facet on the slicer. [*middle left*] layout of 4 mas plate scale lenslet pre-optics; [*middle right*] spot diagram at various field points at the lenslet plane. [*lower left*] layout of 9 mas plate scale lenslet pre-optics; [*lower right*] spot diagram at various field points at the lenslet plane.

Alignment tolerances of the TMA's are eased by allowing compensation with the image plane tilt and with the tilt of the mirror nearest the image. This compensation corrects astigmatism created by other misalignments. If we allow the adjustments the alignment tolerances are eased by approximately a factor of 4. The tolerances, therefore, would be as follows: ~0.1% radius of curvature,~<λ/10 P-V wavefront error @633nm,100-200µ decenters, and 1 arcmin tilts. The compensating tilts would be on the order of 2-3 mrad.

A first cut at the stability requirements assumes that a 1µ movement on the image plane (cf. 15µ pixel) would be the requirement, but the timescale of such movements needs to be established. This requirement would result in stability tolerances of 0.5-1.5µ decenter and 1.5-3.0 µrad tilts. Initial inquiries indicate that the fabrication of the aspheres should not be particularly expensive or problematic.

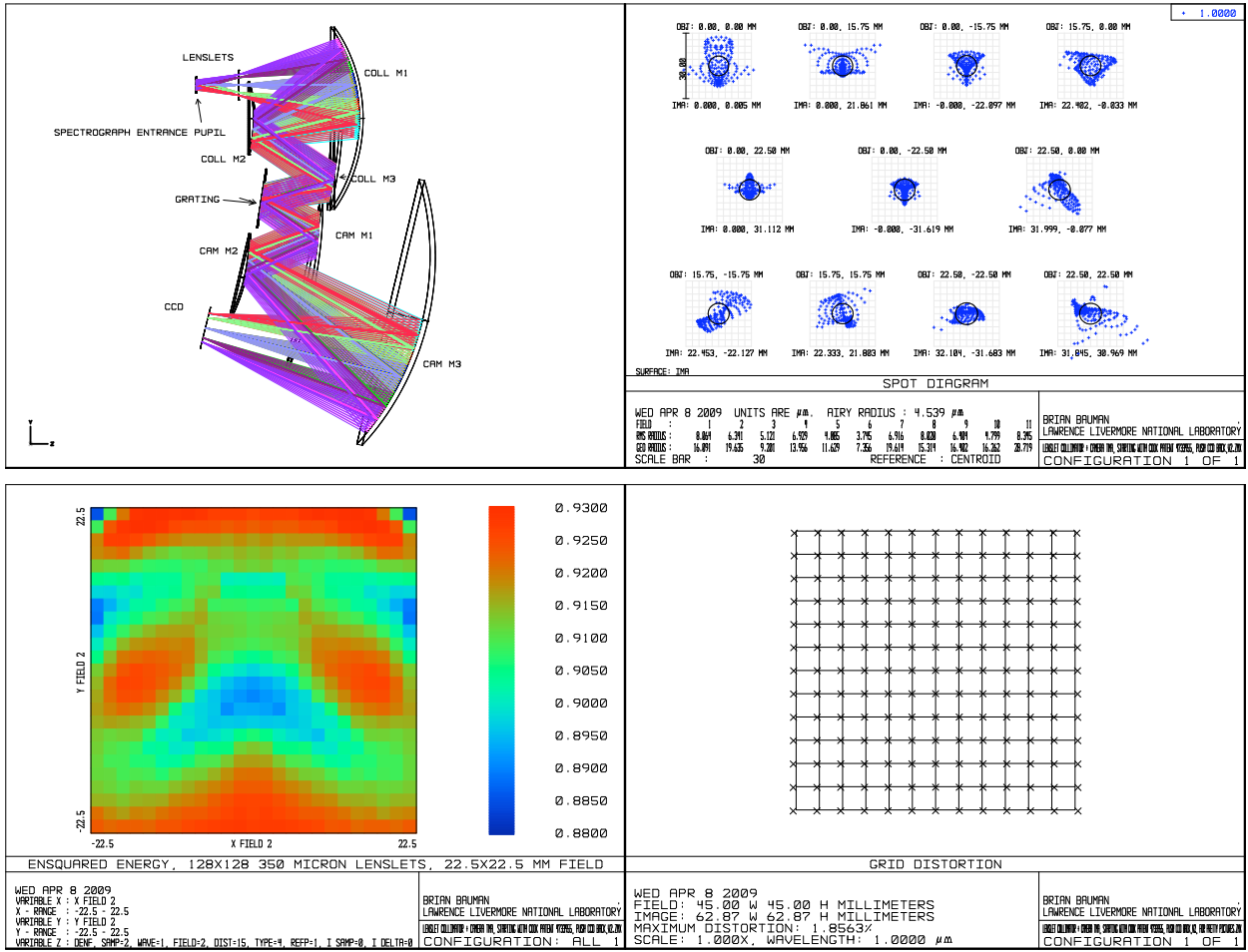

Figure 7: (*upper left*) Layout of lenslet spectrograph collimator and camera TMA, with off-axis parents shown. (*upper right*) spot diagram at the image plane for several positions within the field. The grid size is 2 CCD pixels wide; (*lower left*) diffraction ensquared energy plot, showing that 88-92% of the total energy is captured within 2 pixels; (*lower right*) grid distortion plot, showing a maximum distortion of approximately 1%.

## 7. IMAGE SLICER IFS

The image slicer IFS consists of an 88-channel all-spherical integral field unit, a refractive collimator and onwards to a shared grating turret, TMA camera and 4k x 4k IR detector. The latter three items form part of the lenslet IFS summarized in the previous section.

## 7.1 Image slicer

An image slicer integral field unit (IFU) samples the coarsest IRIS spatial scales and hence the widest fields of view. The image slicer IFU is composed of 88 channels, and is a 3-mirror system per channel. The IRIS image slicer converts the 2-dimensional field of view into 2 monolithic long slits that are optically fed by a single common pupil. The top-level specifications of the image slicer optics are shown in Table 4.

A cartoon version of the IRIS image slicer is shown in Figure 8. It is noted that the diagram is for instructive purposes only and does not accurately represent scale nor ray geometry. Light from the slicer pre-optics enters from the top-left and is reflected by the single slicer mirror that is flat (zero power). A real image of the slicer entrance pupil would form if the slicer mirror were not in place. The slicer entrance pupil is formed on or close to the slicer mirror 2. This mirror, referred to as the "pupil mirror" directs the light to the slicer mirror 3 and together they form the slitlet image. Both mirrors 2 and 3 are powered. Slicer mirror 3, in addition, forms an image of the slicer entrance pupil such that it is correctly located and coincident with the other 87 channels. This detail is not shown.

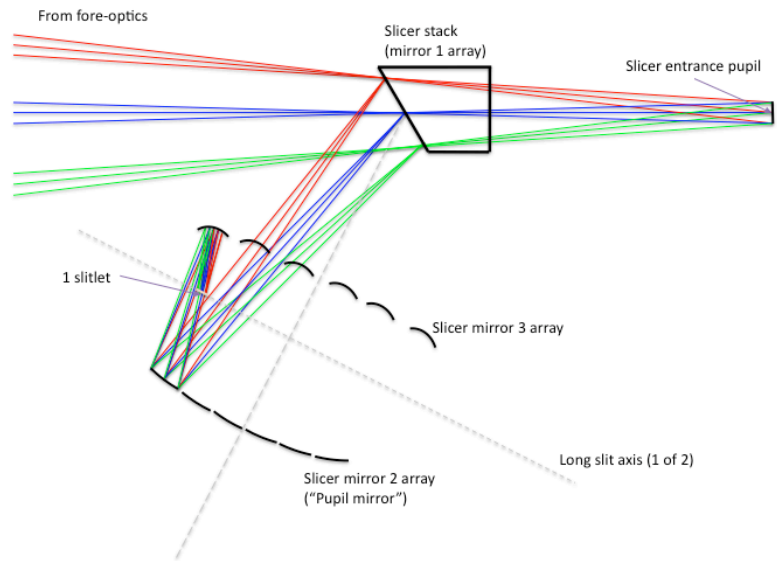

Figure 8: A schematic layout only of the IRIS image slicer IFU and formation of the long slit. Rays for a single channel are shown out of the 88 in total. The layout is not to scale and is for clarification only. See text for details.

**De-magnification**
There are 88 slicer mirrors in the slicer stack, each spanning 45 spaxels. Neglecting inter-slitlet gaps for the purposes of this analysis, this results in a total length of 1980 spaxels for each of the long slits. Without de-magnification this is equivalent to a physical dimension of 1980 x 0.4mm or 792mm. Such a size requires a collimator first element of ~1m. As such, the slicer stack field is reduced in size by a factor of 6, from 18mm at the slicer stack to ~3mm at the long slit. With a close packed array of pupil mirrors this results in a pupil size of ~6mm, a manageable size for polished glass mirrors.

**Varying magnification**
The use of 2 powered mirrors to re-image both the slicer field to the slitlet and the entrance pupil to the exit pupil results in a slightly different magnification across the field. This difference arises from the need for slicer mirror 2, the pupil mirror, to shift slightly along the z-axis and hence move away from the entrance pupil to ensure all 88 slitlet images form in two parallel lines on a single plane. The effect is ~8% and considered acceptable by the IRIS science team.

**Two long slits**
To optimally utilize the square 4k x 4k detector, the IRIS image slicer reformats the pre-optics focal plane to form 2 (real) long slits that are imaged onto the detector. In turn, each long slit consists of alternating slit images, one for each channel, that for mirror packing reasons are ½ a slit length in separation.

Table 4: First order specifications for the image slicer mirrors

|  | Radius of Curvature (mm) | Conic constant | Clear aperture Size (mm) |
|---|---|---|---|
| Slicer mirror | Infinite | - | 0.4 x 35.2 |
| Slicer mirror 2 (Pupil mirror) | 144.3 (concave) | 0 | 6.3 diameter |
| Slicer mirror 3 (Slit mirror) | 107.8 (concave) | 0 | ~3 x 6 rectangular |

**Anamorphic magnification**

The baseline design for the IRIS image slicer has no anamorphic magnification. Anamorphic magnification, that is to say the uneven magnification of the field along 2 perpendicular axes, is to enable Nyquist sampling of the spectral resolution peak without an associated increase in the magnification of the image in the spatial direction. Therefore, the uneven magnification, in actuality performed in the image slicer fore-optics, results in a field that is 2:1 spectral to spatial direction. For certain types of science, namely bright target work, this can result in a roughly sqrt(2) gain in the SNR. Anamorphic magnification is unique to image slicer IFUs.

For the 50mas spaxel scale, if anamorphic magnification were to be adopted, the resulting spectrograph camera would need to be twice as fast in the direction parallel to the long slit as it is currently, or roughly F/2.1. This would have made the camera very challenging. More importantly, the symmetry between the fastest required lenslet camera design and that of the, non-anamorphically magnified, 50mas slicer scale (F/4.2) enables direct sharing of one of the most expensive assembles in the IRIS instrument, namely the camera TMA. This was evaluated as a large incentive for keeping the non-anamorphically magnified design for the 50mas scale.

We are, however, at liberty to design the 25mas image slicer fore-optics to produce a 2:1 magnification and use the same camera TMA. This would result in a field that is twice as large in the spatial direction but remains the same size in the spectral direction. Similarly as before, we decided to pursue the simplest fore-optics for the concept design study, which does not result in an anamorphically magnified field for either the 25mas or 50mas fields. Investigation into an alternative solution, dictated by the benefit to science observations, is not ruled out for the early Preliminary design phase.

**Image slicer summary**

An in depth analysis of the image slicer is not possible here, however, we present the salient points:

- The IRIS image slicer demagnifies the input image and contains flat or spherical mirrors only. There are 3 mirrors per channel;

- There are 88 channels arranged in two levels of 44 channels, each channel providing a slitlet of length 45 spaxels. The total number of spaxels utilized is therefore ~97% of the detector area (3960/4096);

- The imaging and encircled energy performance across the field is excellent (encircled energy far exceeds >80% in 2x2 detector pixels);

- There is a plate scale change from the center of the long slit imaged onto the detector to the edge of value ~8%;

- The slicer unit will likely be made of low expansion glass rather than diamond turned Aluminum given the design is all-spherical– the technology exists to make this IFU today.

### 7.2 Slicer collimator

The slicer collimator was chosen to be refractive for packaging and performance reasons. The current design is a 3-glass air-spaced triplet with barium fluoride, fused silica, and zinc selenide. The slicer collimator is likely overspecified with respect to its chromatic performance: axial color was fully penalized. When the chromatic requirements are revisited in the Preliminary design phase, it is expected that a two-glass design will result. All surfaces are spherical and aspherizing would not help as the dominant aberrations are chromatic. When lateral color is neglected, the rms spot size would be less than 10 microns at the slit mirror plane. Note that lateral color is common when the lens is distant from the pupil in this manner. The effect, however, is judged to be small, since it only translates the spectra on the CCD.

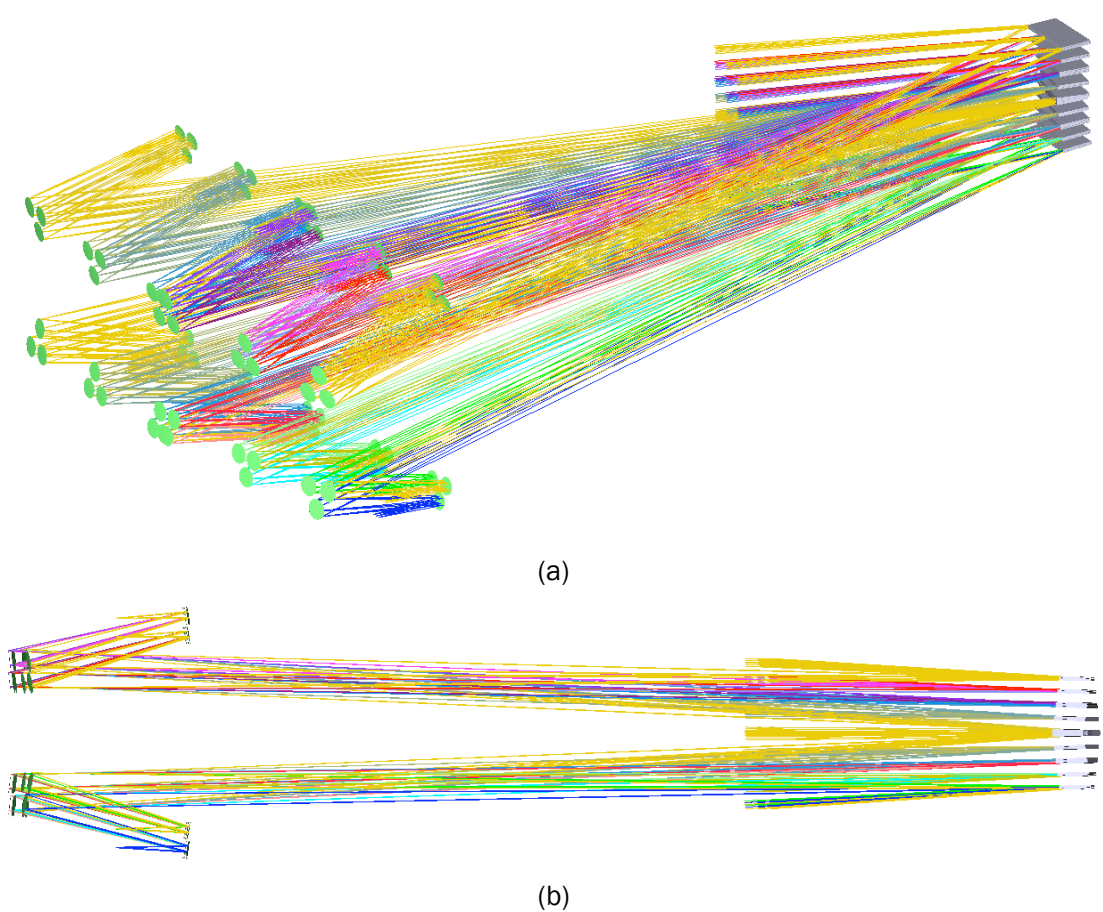

Figure 9: Zemax raytraces of 36 of the 88 channels constituting the IRIS image slicer shown in (a) Isometric and; (b) Side view. The trace begins 100mm prior to the slicer stack and ends at the slicer long slit plane.

The slicer collimator triplet is approximately 250mm in diameter, which is largely a result of the required 500 mm stand-off distance between the diffraction grating and the lens (i.e., the "eye relief").

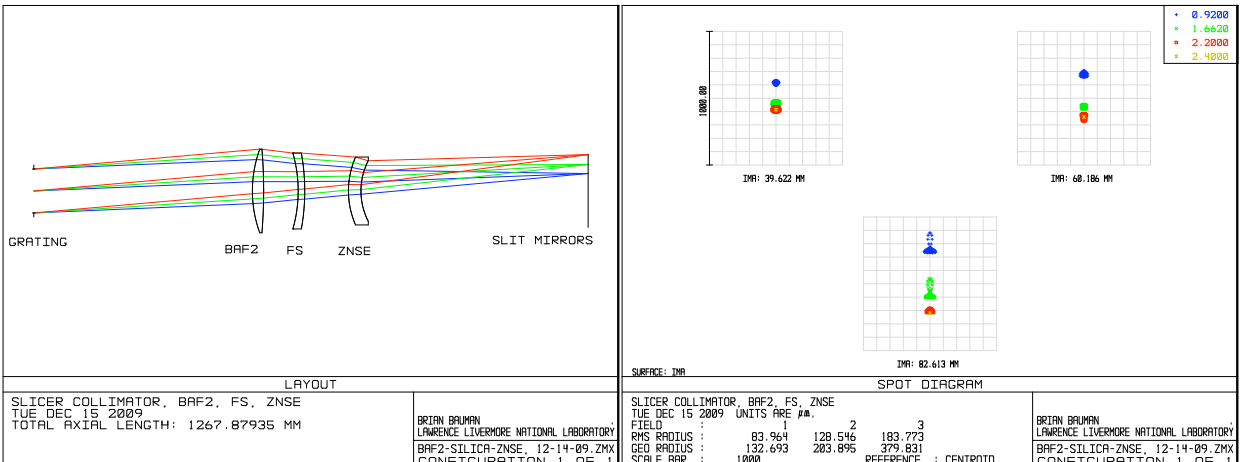

Figure 10: (*left*) Layout of the refractive slicer collimator, from grating to slit mirrors (light travels from right to left in the picture). (*right*) Spot diagrams showing mostly lateral color present in the design.

# 8. FILTERS, GRATINGS, LYOT MASK AND ADC

## 8.1 Filters and gratings

The IRIS IFS observing modes are summarized in Table 5. The 22 or so minimum number of filters required for the IFS observing modes are stored in a 3-level filter wheel located after the IFS common collimator. The filters are small, roughly ½-1" in diameter.

Table 5: Observing modes for the IRIS IFS

| # | Grating groove Den (mm$^{-1}$) | Blaze angle | Filters | Field mode | Resolution | Narrow band Filter Set | | | |
|---|---|---|---|---|---|---|---|---|---|
| | | | | | | Filter | Central (μm) | Min (μm) | Max (μm) |
| 1 | 341 | 9.86 | Zbb | Slicer | 4000 | ZN1 | 0.8613 | 0.8400 | 0.8830 |
| | | | Zn1-Zn4 | Lenslet 112x128 | 4000 | ZN2 | 0.8963 | 0.8742 | 0.9190 |
| 2 | 290 | 9.86 | Ybb | Slicer | 4000 | ZN3 | 0.9328 | 0.9098 | 0.9564 |
| | | | Yn1-Yn4 | Lenslet 112x128 | 4000 | ZN4 | 0.9708 | 0.9469 | 0.9954 |
| 3 | 249 | 9.86 | Jbb | Slicer | 4000 | YN1 | 1.0104 | 0.9854 | 1.0359 |
| | | | Jn1-Jn4 | Lenslet 112x128 | 4000 | YN2 | 1.0515 | 1.0256 | 1.0781 |
| 4 | 194 | 9.86 | Hbb | Slicer | 4000 | YN3 | 1.0943 | 1.0673 | 1.1220 |
| | | | Hn1-Hn5 | Lenslet 112x128 | 4000 | YN4 | 1.1389 | 1.1108 | 1.1677 |
| 5 | 145 | 9.86 | Kbb | Slicer | 4000 | JN1 | 1.1853 | 1.1560 | 1.2153 |
| | | | Kn1-Kn5 | Lenslet 112x128 | 4000 | JN2 | 1.2336 | 1.2031 | 1.2648 |
| 6 | 161 | 9.86 | H+K | Lenslet 16x128 | 4000 | JN3 | 1.2838 | 1.2521 | 1.3163 |
| 7 | 709 | 20.54 | Zn1-Zn4 | Slicer | 8000 | JN4 | 1.3361 | 1.3031 | 1.3699 |
| | | | Zbb | Lenslet 16x128 | 8000 | HN1 | 1.5113 | 1.4740 | 1.5495 |
| 8 | 606 | 20.54 | Yn1-Yn4 | Slicer | 8000 | HN2 | 1.5728 | 1.5340 | 1.6126 |
| | | | Ybb | Lenslet 16x128 | 8000 | HN3 | 1.6369 | 1.5965 | 1.6783 |
| 9 | 516 | 20.54 | Jn1-Jn4 | Slicer | 8000 | HN4 | 1.7036 | 1.6615 | 1.7467 |
| | | | Jbb | Lenslet 16x128 | 8000 | HN5 | 1.7730 | 1.7292 | 1.8178 |
| 10 | 398 | 20.54 | Hn1-Hn5 | Slicer | 8000 | KN1 | 2.0096 | 1.9600 | 2.0604 |
| | | | Hbb | Lenslet 16x128 | 8000 | KN2 | 2.0914 | 2.0398 | 2.1443 |
| 11 | 310 | 20.54 | Kn1-Kn3 | Slicer | 8000 | KN3 | 2.1766 | 2.1229 | 2.2317 |
| 12 | 282 | 20.54 | Kn4-Kn5 | Slicer | 8000 | KN4 | 2.2653 | 2.2094 | 2.3226 |
| 13 | 296 | 20.54 | Kbb | Lenslet 16x128 | 8000 | KN5 | 2.3575 | 2.2993 | 2.4172 |
| 14 | 872 | 26.0 | Zbb | Lenslet 16x128 | 10,000 | Broadband filter set | | | |
| 15 | 741 | 26.0 | Ybb | Lenslet 16x128 | 10,000 | Z | 0.928 | 0.840 | 1.026 |
| 16 | 637 | 26.0 | Jbb | Lenslet 16x128 | 10,000 | Y | 1.092 | 0.988 | 1.206 |
| 17 | 497 | 26.0 | Hbb | Lenslet 16x128 | 10,000 | J | 1.270 | 1.149 | 1.403 |
| 18 | 368 | 26.0 | Kbb | Lenslet 16x128 | 10,000 | H | 1.629 | 1.474 | 1.800 |
| 20 | Mirror | 22.5 | Any | Any | NA | K | 2.182 | 1.975 | 2.412 |

## 8.2 Lyot stop

Precise rejection of thermal background is accomplished using a lyot mask placed on a rotating stage at the reimaged pupil location. The mask itself is small, however, the shape must fully match the TMT pupil and rotate precisely to accomplish the required thermal background requirement.

## 8.3 IFS ADC

The Atmospheric dispersion corrector for the IRIS IFS is located after the rotating lyot mask and close to the filter wheels. The ADC is a crossed amici design based on the larger versions required for the IRIS imager ADC. Further details can be found in Phillips et al [9].

# 9. MECHANICAL DESIGN

An in depth presentation of the IRIS IFS mechanical design is outside the realm of this paper. The mechanical design has been taken to concept design level and beyond in cases where direct copy of previous hardware is possible (eg MOSFIRE [10]). The IFS risk items and mechanisms are briefly addressed below.

## 9.1 Risk items

Two hardware items were identified as moderate risk areas during the concept design: the double TMA assembly and grating turret. Both will be prototyped in the next phase of funding.

## 9.2 Mechanisms

There are a total of 8 mechanisms located inside the cryogenic dewar that belong to the IFS: Lyot stop; Filter wheel, ADC, 4mas/9mas exchanger, 25mas/50mas exchanger, fold mirror, grating turret and detector focus. All linear slides are based on the successful 2-position mechanism developed for MOSFIRE, GPI [11] etc. that provide high accuracy and are proven technology. The grating turret is by far the riskiest mechanism. This is because we require exceptional stability of the spectra at the detector from a mechanism that contains as many as 15 reflective gratings. Because of this, the mechanism will be prototyped in the next phase. A small selection of further mechanisms is shown in Figure 11.

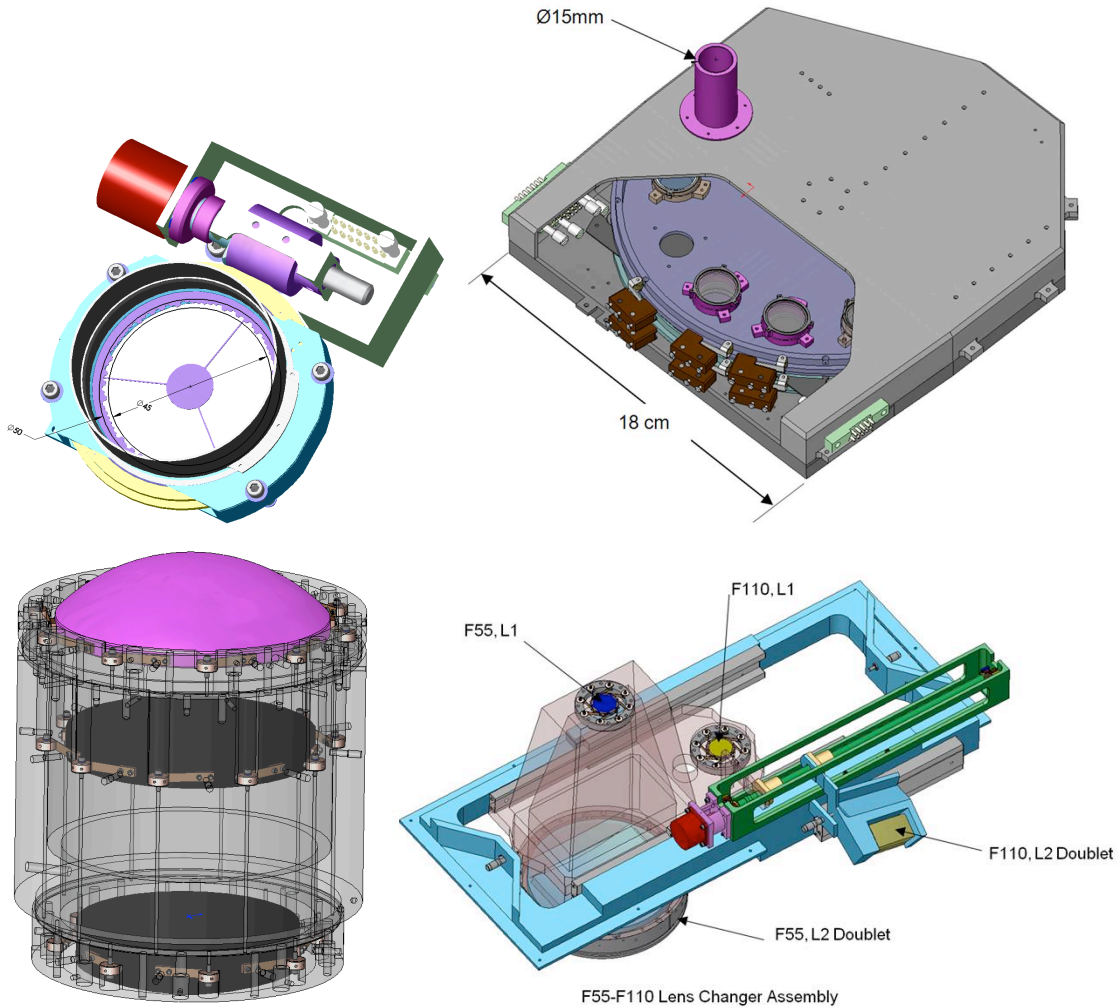

Figure 11: [*upper left*] the precision rotating Lyot mask; [*upper right*] the IFS filter wheel; [*lower left*] the image slicer collimator and; [*lower right*] the 25mas/50mas exchanger mechanism.

## ACKNOWLEDGEMENTS

The authors gratefully acknowledge the support of the TMT partner institutions. They are the Association of Canadian Universities for Research in Astronomy (ACURA), the California Institute of Technology and the University of California. This work was supported as well by the Gordon and Betty Moore Foundation, the Canada Foundation for Innovation, the Ontario Ministry of Research and Innovation, the National Research Council of Canada, the Natural Sciences and Engineering Research Council of Canada, the British Columbia Knowledge Development Fund, the National Astronomical Observatory of Japan (NAOJ), the Association of Universities for Research in Astronomy (AURA) and the U.S. National Science Foundation.